\documentclass{article}
\usepackage{graphics} 
\usepackage{graphicx} 
\usepackage{hyperref} 
\usepackage{subcaption}
\usepackage{multirow} 
\usepackage{tabularx} 
\usepackage{color} 
\usepackage{url}
\usepackage{textcomp} 
\usepackage{amsmath} 
\usepackage{amssymb} 
\usepackage{amsfonts} 
\usepackage[margin=1in]{geometry}
\usepackage[numbers]{natbib}
\usepackage[linesnumbered,ruled]{algorithm2e}

\definecolor{color-1}{rgb}{1,1,1}

\title{A History Equivalence Algorithm for Dynamic Process Migration}

\author {Gargi Bakshi\footnote{Gargi Bakshi is a final year undergraduate student of Computer Science \& Engineering at IIT Bombay.}\\ \small{Department of Computer Science \& Engineering,}\\ \small {Indian Institute of Technology Bombay} \\ \small{Mumbai, India - 400076.}
\and
Rushikesh K. Joshi \footnote{Rushikesh K. Joshi is a Professor of Computer Science \& Engineering at IIT Bombay. Email for Correspondence: rkj@iitb.ac.in} \\ \small{Department of Computer Science \& Engineering,} \\ \small{Indian Institute of Technology Bombay}\\ \small{Mumbai, India - 400076.}}

\begin{document}
\date{}
\maketitle
\begin{abstract}
Dynamic changes in processes necessitate the notion of state equivalence between the old and new workflows. In several cases, the history of the workflow to be migrated provides sufficient context for a meaningful migration. In this paper, we present an algorithm to find the equivalence mapping for states from the old workflow to the new one using a trail-based consistency model called history equivalence. The algorithm finds history equivalent mappings for all migratable states in the reachability graph of the process under migration. 
It also reports all non-migratable states that fall in the change region for a given pair of old and new Petri Nets. The paper presents the algorithm, its working, and an intuitive proof. The working is demonstrated through a couple of illustrations.
\end{abstract}


\maketitle

\section{Introduction and Related Literature}
Business processes often evolve, which requires the underlying structures representing the processes to change. While migrating an instance from an old process to a new one, it is crucial to find suitable equivalent states to minimize the disruption caused by the process migration. The notion of migration in Petri net-based processes was initially explored by Ellis et al. \cite{dynamic_change_within_wf_systems}, which was followed by several other works including \cite{supporting_adhoc_changes_in_distributed_wf_systems} and \cite{workflow_evolution}. The notion of consistency in process migration was subsequently unified through the taxonomy defined by Pradhan and Joshi \cite{a_taxonomy}, in which several new consistency notions have also been defined. In this paper, {\em history-based consistency}, one of the nine broad consistency models defined in \cite{a_taxonomy} is being explored for automation of process migration.

For this consistency criterion, an algorithm is presented to find the mapping of migratable states in the reachability graph from a net under migration to the new net. Our algorithm also reports states that fall under change region. This notion of change region was introduced and explored in the early works of Ellis et al.\cite{dynamic_change_within_wf_systems} and Van der Aalst \cite{vda_changeregion}. The actual set of states in the change region depends on the consistency criterion used for migration. 

The algorithm uses a Workflow net model for representing processes, with each business state being defined by a state consisting of one or more {\em places}, and each task being defined by a {\em transition}. Workflow nets have been modelled in Petri Nets, and one of the early uses of Petri nets for workflow nets can be explored through the work of Van der Aalst \cite{WFnet_def}, \cite{van1998three}. The history-based consistency model called {\em history equivalence} focuses on the past transition firings that have led to a state to define equivalence. We present an algorithm to obtain the history-based equivalence states of a dynamically migrated process.

\section{Organization of the Paper}
 The next section provides definitions of workflow nets, consistency, and trail-based consistency models, which include the history equivalence consistency. Section \ref{ourproblem} describes the problem statement and the motivation behind the algorithm. Section \ref{for_algo} gives a detailed description of the algorithm and all its supporting components with the help of a few examples. Section \ref{intuition} provides an intuitive proof for the correctness of the algorithm.

\section{Definitions} \label{defs}
\subsection{Workflow Nets}
A Workflow Net (WF-net) \cite{WFnet_def} is a type of Petri Net \cite{murata_pn} designed for modelling and analyzing workflows, featuring control flow patterns like Sequence, Choice, Concurrency, and Loops. A WF-net $N$ can be defined by a tuple $(P, T, F)$, where $P$ is a finite set of places, $T$ is a finite set of transitions, and $F \subseteq (P \times T) \cup (T \times P)$ is a finite set of arcs. Some properties of WF-nets are listed below. \vspace{-2pt}
\begin{itemize}
    \item Unique source {\em init} and unique sink {\em end}. \vspace{-5pt}
    \begin{itemize}
        \item The initial marking $M_i$ is unique, and it is given by one token in {\em init}. \vspace{-2pt}
        \item The terminal marking $M_e$ is unique, and it is given by one token in {\em end}. \vspace{-3pt}
    \end{itemize}    \item All places and transitions except {\em init} and {\em end} appear in at least one directed path from {\em init} to {\em end}. \vspace{-5pt}
    \begin{itemize}
        \item Every transition can be fired from some marking $M$ reachable from $M_i$.  \vspace{-2pt}
        \item Every marking $M$ reachable from $M_i$ eventually reaches $M_e$.  \vspace{-3pt}
    \end{itemize}
\end{itemize} \par
\begin{figure}
    \centering
    \begin{subfigure}[b]{0.4\textwidth}
        \centering
        \includegraphics[width=\textwidth]{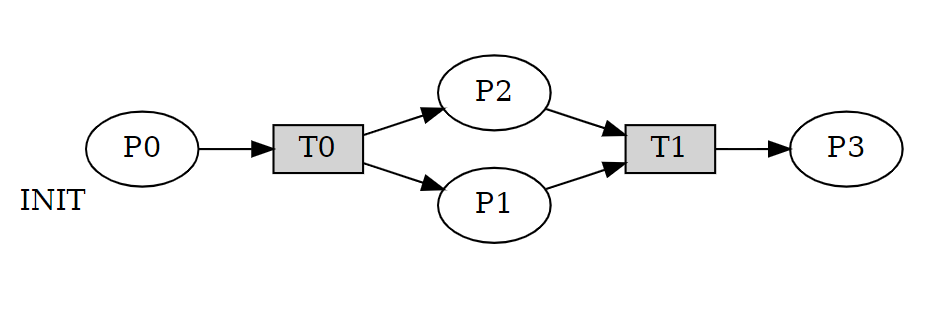}
        \caption{a valid WF-net which satisfies all the mentioned properties}
        \label{fig:validWFnet}
    \end{subfigure}
    \hfill
    \begin{subfigure}[b]{0.4\textwidth}
        \centering
        \includegraphics[width=\textwidth]{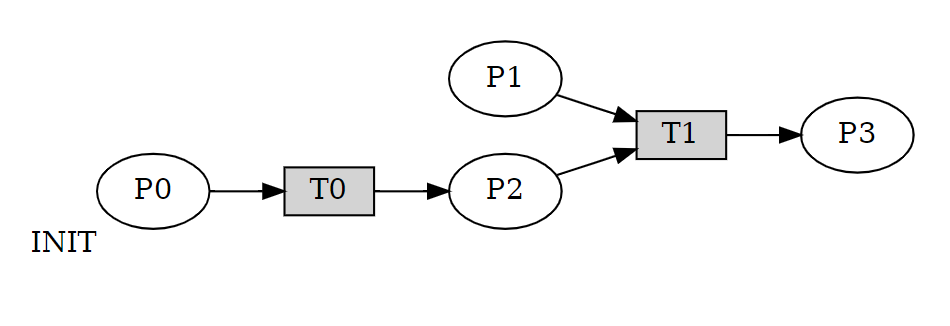}
        \caption{not a WF-net because place $\text{P}_1$ does not appear in a directed path from the net's {\em init} to {\em end}}
        \label{fig:invalidWFnet}
    \end{subfigure}
    \caption{A Petri Net that is a WF-net and a Petri Net that isn't}
    \label{fig:main1}
\end{figure}

Figure \ref{fig:validWFnet} is a valid workflow net. Figure \ref{fig:invalidWFnet} is not a valid workflow net because place $\text{P}_1$ does not appear in a directed path. A workflow net can have loops. The nets are 1-bounded nets, with at most one token per place. Hence, they generate a finite reachability graph, which makes it possible for us to find the equivalence mapping for all migratable states.

\subsection{Consistency in Dynamic Migration} 
A dynamic change in a process refers to a modification in its workflow net. An instance of the old workflow net can be migrated to the new workflow net based on some equivalence criteria. The problem of consistency deals with identifying equivalent instances to facilitate migration from old states to new states. 
\par
We can have several equivalence criteria and decide which ones are relevant for a particular migration based on the semantics of the specific workflow application being modelled \cite{10.1007/3-540-44895-0_4}. A taxonomy of consistency models to establish equivalence between two control-flow states of two given nets is described in \cite{a_taxonomy}. It introduces various models that use different parameters like relative time frames, traces and net structure to establish equivalence. In this paper, we focus on \emph{history equivalence}, which is one of the trail-based consistency models, as defined in \cite{a_taxonomy}.


\subsection{Trail-Based Consistency Models}
A trace is the firing sequence from the initial marking to another. A trail is the exact trace of the execution of an instance reaching a given current state that is under migration.\par
Trail-based consistency models define equivalence based on the past transition firings that have led to the current state. \par
Let the old net be $N (P, T, F)$ and the new net $N^\prime (P^\prime, T^\prime, F^\prime)$ with initial markings $M_i$ and ${M_i}^\prime$ respectively. Considering markings $M$ and $M^\prime$ that can be obtained through traces $\sigma$ and $\sigma^\prime$ in the old and new net respectively, the equivalence of markings $M$ and $M^\prime$ can be established based on four types of trail-based consistency models \cite{a_taxonomy}:
\vspace{-5pt}
\begin{itemize}
    \item Trace Equivalence: $\sigma = \sigma^\prime$ 
    \item History Equivalence: $s(\sigma) = s(\sigma^\prime)$ 
    \item Purged Trace Equivalence: $\sigma|_{T \cap T^\prime} = \sigma^\prime|_{T \cap T^\prime}$ 
    \item Purged History Equivalence: $s(\sigma|_{T \cap T^\prime}) = s(\sigma^\prime|_{T \cap T^\prime})$ 
\end{itemize} 
\vspace{-5pt}
$s(\sigma)$ denotes the set of transitions in the trace $\sigma$, and $\sigma|_{T \cap T^\prime}$ denotes the subsequence of transitions in the trace $\sigma$ that are also present in $T^\prime$. Purged equivalences can be used when $\sigma$ and $\sigma^\prime$ have one or more transitions in common.
History equivalence uses trace sets to decide if two states are equivalent or not. We describe an algorithm that uses this model to find the equivalence mapping from an old WF-net to the new WF-net. To establish migration equivalence of two states from two processes, an equivalent $(\sigma, \sigma^\prime)$ pair needs to be found, showing equivalence over the chosen equivalence criterion.

\section{Our Problem} \label{ourproblem}
Several processes involve tasks that can be performed in more than one chronological order without affecting the expected outcome. When such processes migrate to new processes, equivalence criteria that do not take into account the order of task execution are good candidates for establishing state equivalence from the old to the new process. The reason is that, for such processes, one does not need to use the chronology.

For dynamic migration tasks, it is beneficial to have an algorithm to map states in the old process to equivalent states in the new process as per the consistency criterion selected. History equivalence is a consistency model that is well-suited for migrations that do not need to keep track of chronology, since it only needs information about task completion status. We now develop an algorithm to obtain a history equivalence-based mapping between states in the old and new workflow nets.

\section{Equivalence-Mapping Algorithm for History Equivalence} \label{for_algo}
The algorithm to find the history equivalence mapping between two nets (given by Algorithm~\ref{algo:find_equivalence_mapping}) has three main steps: \vspace{-3pt}
\begin{enumerate}
    \item Build reachability graphs for the old and the new net. The reachability graph of a Petri net is a directed graph, $G = (V, E)$, where each node, $v \in V$, represents a reachable marking and each edge, $e \in E$, represents a transition between two reachable markings. Figure  \ref{fig:pn}  shows a workflow net, and its reachability graph is shown in Figure \ref{fig:rg}.

\begin{figure}
    \centering
    \begin{subfigure}[b]{0.4\textwidth}
        \centering
        \includegraphics[width=\textwidth]{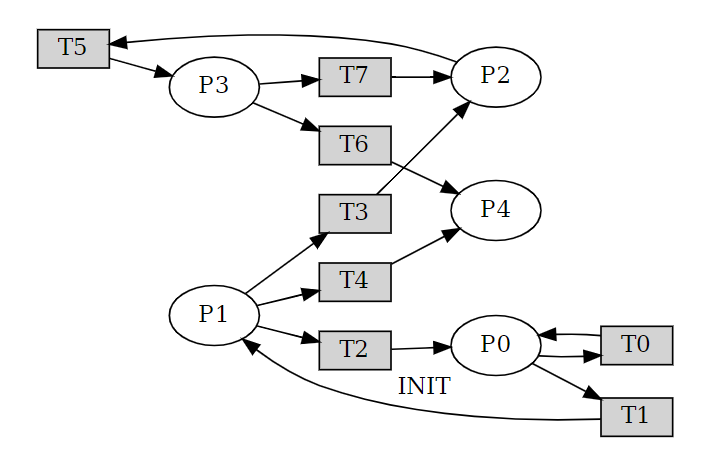}
        \caption{WF-net}
        \label{fig:pn}
    \end{subfigure}
    \hfill
    \begin{subfigure}[b]{0.4\textwidth}
        \centering
        \includegraphics[width=\textwidth]{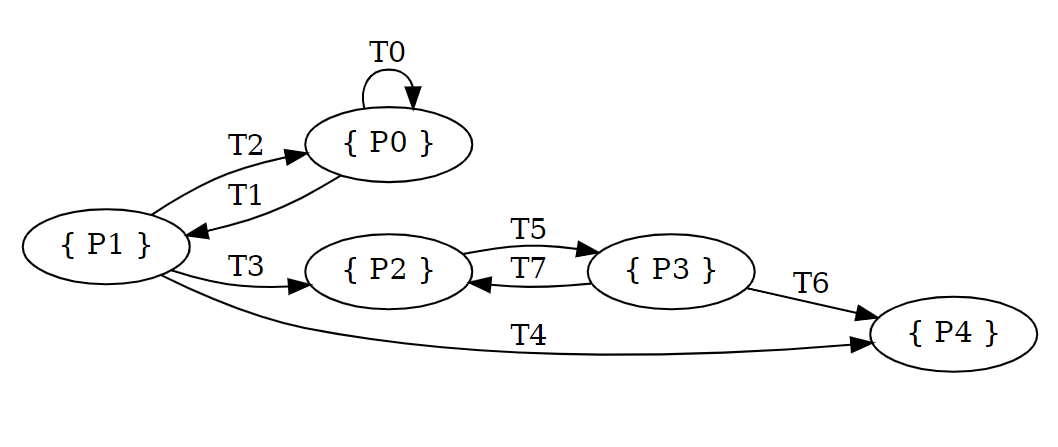}
        \caption{Reachability graph}
        \label{fig:rg}
    \end{subfigure}
    \caption{WF-net (a) and its reachability graph (b)}
    \label{fig:main}
\end{figure}

    \item For each node $N$ (a node represents a state, which is a reachable marking in the net) in the reachability graphs of the old and the new nets, find all unique sets of transitions, each corresponding to one or more possible traces that lead to $N$. These are called Trace Transitions Sets (TTSs), which are notably not the set of traces themselves as the example below illustrates. It can be noted that for a given state $N$, the number of TTSs is finite, and each TTS is also finite. This is because there are only finitely many transitions in a given net, and the number of TTSs is bounded by the power set, while the cardinality of every TTS is bounded by the number of transitions. However, the traces of unbounded length may be possible due to loops. In the case of history equivalence, our interest is in TTSs. The crux of the algorithm lies in finding the TTSs.

\begin{figure}[htbp]
    \centering
    \includegraphics[width=0.8\textwidth]{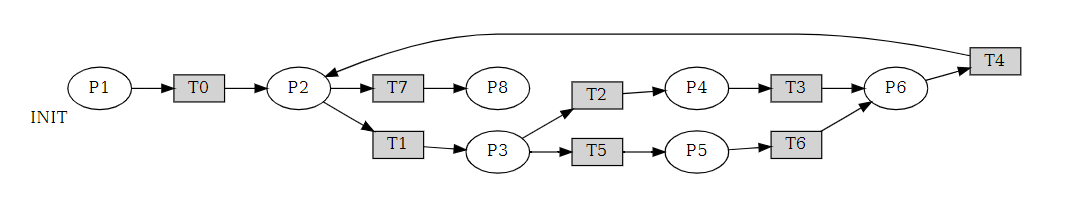}
    \caption{WF-net with cycles}
    \label{fig:complex_paths_pn}
\end{figure}

\begin{figure}[htbp]
    \centering
    \includegraphics[width=0.6\textwidth]{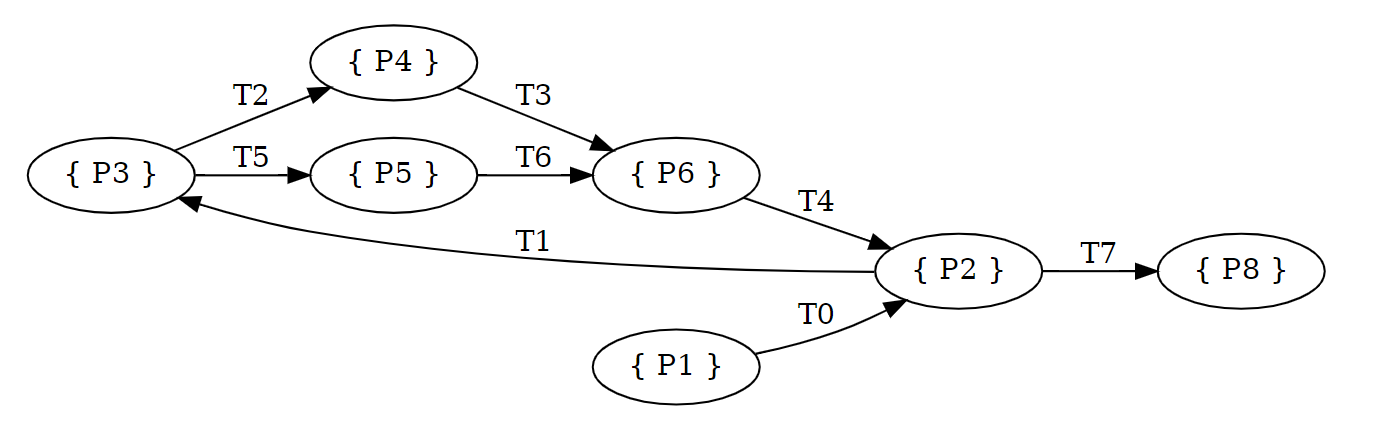}
    \caption{Reachability Graph with loops, with initial state as $\{P_1\}$}
    \label{fig:complex_paths_rg}
\end{figure}

Figure \ref{fig:complex_paths_rg} shows the reachability graph for the WF-net given in Figure \ref{fig:complex_paths_pn}. For the marking $\{\text{P}_2\}$, the trace transition sets are:
\begin{align*}
    &\{T_0\} \\
    &\{T_0, T_1, T_2, T_3, T_4\} \\
    &\{T_0, T_1, T_5, T_6, T_4\} \\
    &\{T_0, T_1, T_2, T_5, T_3, T_6, T_4\}
\end{align*}

It can be noted that the given reachability graph has an infinite number of traces corresponding to marking $\{\text{P}_2\}$ from initial state $\{\text{P}_1\}$. However, the TTS corresponding to $\{\text{P}_2\}$ is finite. The set of transitions for each of these infinite traces is exactly one of the above four TTSs. For example, the trace $<T_0 \hspace{-0.7mm} \rightarrow \hspace{-0.7mm} T_1 \hspace{-0.7mm} \rightarrow \hspace{-0.7mm} T_5 \hspace{-0.7mm} \rightarrow \hspace{-0.7mm} T_6 \hspace{-0.7mm} \rightarrow \hspace{-0.7mm} T_4 \hspace{-0.7mm} \rightarrow \hspace{-0.7mm} T_1 \hspace{-0.7mm}\rightarrow \hspace{-0.7mm} T_2 \hspace{-0.7mm} \rightarrow \hspace{-0.7mm} T_3  \hspace{-0.7mm} \rightarrow \hspace{-0.7mm} T_4>$ has $\{T_0, T_1, T_2, T_5, T_3, T_6, T_4\}$ as its TTS, and trace $<T_0>$ has $\{T_0\}$ as its TTS. Both these traces are possible given marking $\{\text{P}_2\}$. Similarly, trace $<T_0 \hspace{-0.7mm}\rightarrow \hspace{-0.7mm} T_1 \hspace{-0.7mm}\rightarrow \hspace{-0.7mm}T_5\hspace{-0.7mm} \rightarrow \hspace{-0.7mm}T_6 \hspace{-0.7mm}\rightarrow\hspace{-0.7mm} T_4>$ and trace $<T_0 \hspace{-0.7mm}\rightarrow\hspace{-0.7mm} T_1\hspace{-0.7mm} \rightarrow \hspace{-0.7mm}T_5 \hspace{-0.7mm}\rightarrow\hspace{-0.7mm} T_6 \hspace{-0.7mm}\rightarrow\hspace{-0.7mm} T_4\hspace{-0.7mm} \rightarrow \hspace{-0.7mm}T_1\hspace{-0.7mm} \rightarrow \hspace{-0.7mm}T_5 \hspace{-0.7mm}\rightarrow\hspace{-0.7mm} T_6\hspace{-0.7mm} \rightarrow\hspace{-0.7mm} T_4>$ have the same TTS which is $\{T_0, T_1, T_5, T_6, T_4\}$.    

    \item For each of the nodes $N$ in the old net's reachability graph $RG_{\text{old}}$, find the set of nodes in the new net's reachability graph $RG_{\text{new}}$ that have in common one of the TTSs found for $N$ in step 2. If this set is empty for a given node $N$, then $N$ is in the change region. Otherwise, any one of the equivalent nodes found is a possible mapping.
\end{enumerate}

\subsection*{Computing the Trace Transitions Sets}

To find the sets of all possible traces leading to a marking, we need to consider all paths including cycles present in the reachability graph from the initial state to the state under consideration. Therefore, finding of all paths is broken into two steps: finding all "simple" paths (acyclic seed paths) that do not involve loops, and then adding all edges (transitions) present in a cycle that includes a node along a path.\par

The simple paths are found recursively for each node in the graph using the procedure outlined in Algorithm \ref{algo:find_simple_paths}. We then find all the cycles in the reachability graph which are used in making the complete list of trace sets. This can be done using a depth-first search. Cycles found are then used in Algorithm \ref{algo:find_complex_paths}, along with seed paths found in Algorithm \ref{algo:find_simple_paths}. Algorithm \ref{algo:find_complex_paths} then proceeds with each simple TTS and keeps adding each cycle that has a common section with the trace. This procedure is repeated recursively till the entire set of TTSs for that node is found.

For example, in the reachability graph shown in Figure \ref{fig:complex_paths_rg}, the set of simple (seed) traces for $\{\text{P}_2\}$ is $\{< T_0 >\}$, and therefore $\{ T_0 \}$ is the only seed TTS. The cycles found in the graph are $C_1 = (T_1 \hspace{-0.7mm} \rightarrow \hspace{-0.7mm} T_5 \hspace{-0.7mm} \rightarrow \hspace{-0.7mm} T_6 \hspace{-0.7mm} \rightarrow \hspace{-0.7mm} T_4)$ and $C_2 = (  T_1 \hspace{-0.7mm} \rightarrow \hspace{-0.7mm} T_2 \hspace{-0.7mm} \rightarrow \hspace{-0.7mm} T_3 \hspace{-0.7mm} \rightarrow \hspace{-0.7mm} T_4 )$.

Both these cycles can be added along the seed trace  $\{< T_0 >\}$, and so, we add $S_1 = \{ T_0, T_1, T_5, T_6, T_4 \}$ and $S_2 = \{ T_0, T_1, T_2, T_3, T_4 \}$ to the set of TTSs. 
In the next iteration, we make a new TTS by adding the transitions in $C_1$ to $S_2$ (or, alternately, the transitions $C_2$ to $S_1$) to get the set $\{T_0, T_1, T_2, T_5, T_3, T_6, T_4\}$. After this, no further set can be obtained.

\begin{figure}[htbp]
    \centering
    \includegraphics[width=\textwidth]{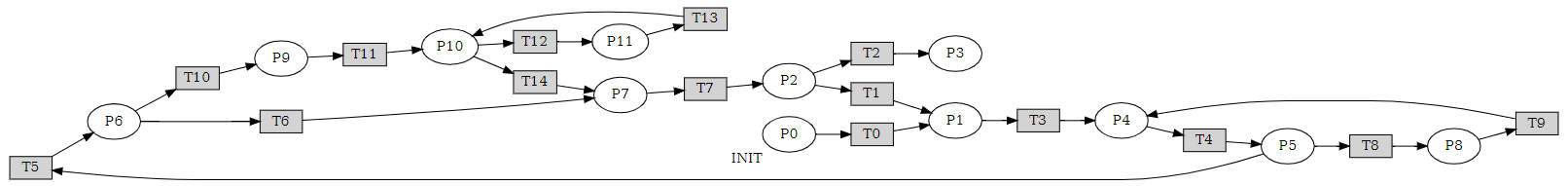}
    \caption{WF-net with many complex paths in the reachability graph (Figure \ref{fig:many_cycles_rg})}
    \label{fig:many_cycles_pn}
\end{figure}

\begin{figure}[htbp]
    \centering
    \includegraphics[width=\textwidth]{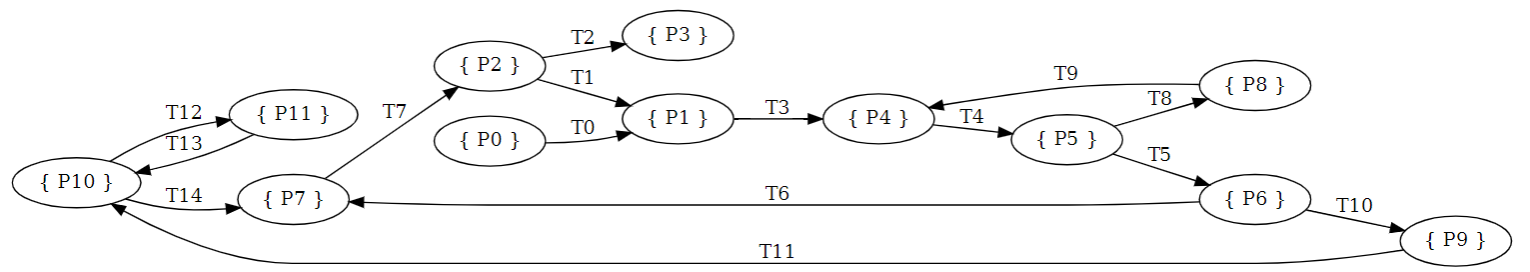}
    \caption{Reachability Graph for WF-net in Figure \ref{fig:many_cycles_pn}}
    \label{fig:many_cycles_rg}
\end{figure}

Now, let us consider another example as given in Figure \ref{fig:many_cycles_pn}. The figure shows a WF-net with many loops. Figure \ref{fig:many_cycles_rg} shows its reachability graph. The following cycles are present in the graph:
\begin{enumerate}
    \item $C_1 = < T_3 \rightarrow T_4 \rightarrow T_5 \rightarrow T_6 \rightarrow T_7 \rightarrow T_1 >$
    \item $C_2 = < T_{12} \rightarrow T_{13} >$
    \item $C_3 = < T_4 \rightarrow T_8 \rightarrow T_9 >$
    \item $C_4 = < T_3 \rightarrow T_4 \rightarrow T_5 \rightarrow T_{10} \rightarrow T_{11} \rightarrow T_{14} \rightarrow T_7 \rightarrow T_1 >$
\end{enumerate}
Let $s(C)$ denote the set of transitions in a cycle $C$. Taking node $\{ P_1 \}$ as an example, we outline below how Algorithm \ref{algo:find_complex_paths} obtains the complete set of TTSs, step-by-step, starting from the simple TTS list $[ \{ T_0 \} ]$. Although we process multiple TTSs together in our analysis, the actual algorithm only considers one set in the list at a time.

In each step, we go through all the elements present in the list of TTSs found from the previous step and, for each TTS, find new cycles that can be added to that TTS. If there is no cycle that can be added for a TTS, or if we have already added the cycles in a previous step, we say that no new cycle can be added, and this is represented by a dash symbol (-). \par
\vspace{5pt} \noindent
\textbf{\textit{Step 1}}
\begin{table}[htbp]
    \centering
    \begin{tabular}{|c|c|}
        \hline
        \textbf{TTS in the list} &\textbf{ New cycles to add} \\
        \hline
        $\{ T_0 \}$ & $C_1, C_4$ \\
        \hline
    \end{tabular}
    \label{tab:step1}
\end{table} \par
This trace finds only two cycles in the graph, which involve at least one node common with the trace.

Resulting TTS list: $[ \{ T_0 \}, \{ T_0, T_1, T_3, T_4, T_5, T_6, T_7 \}, \{ T_0, T_1, T_3, T_4, T_5, T_7, T_{10}, T_{11}, T_{14} \} ]$

\vspace{5pt}\noindent
\textbf{\textit{Step 2}}
\begin{table}[htbp]
    \centering
    \begin{tabular}{|c|c|}
        \hline
        \textbf{TTS in the list} & \textbf{New cycles to add} \\
        \hline
        $\{ T_0 \}$ & - \\
        \hline
        $ \{ T_0 \} \cup s(C_1) \texttt{=} \{ T_0, T_1, T_3, T_4, T_5, T_6, T_7 \}$ & $C_3, C_4$ \\
        \hline
        $ \{ T_0 \} \cup s(C_4) \texttt{=} \{ T_0, T_1, T_3, T_4, T_5, T_7, T_{10}, T_{11}, T_{14} \}$ & $C_2, C_3$ \\
        \hline
    \end{tabular}
    \label{tab:step2}
\end{table} \par
Note that $C_1$ is not a new cycle to add for $\{ T_0 \} \cup s(C_4)$ since $C_4$ is being added to $\{ T_0 \} \cup s(C_1)$. Similarly, since $\{ T_0 \}$ has already been processed, we don't duplicate the cycles that can be added to it. \par
Resulting TTS list: $[ \{ T_0 \}, \{ T_0, T_1, T_3, T_4, T_5, T_6, T_7 \}, \{ T_0, T_1, T_3, T_4, T_5, T_7, T_{10}, T_{11}, T_{14} \}, \\ 
\{ T_0, T_1, T_3, T_4, T_5, T_6, T_7, T_8, T_9 \}, \{ T_0, T_1, T_3, T_4, T_5, T_6, T_7, T_{10}, T_{11}, T_{14} \}, \\\{ T_0, T_1, T_3, T_4, T_5, T_7, T_{10}, T_{11}, T_{12}, T_{13}, T_{14} \}, \{ T_0, T_1, T_3, T_4, T_5, T_7, T_8, T_9, T_{10}, T_{11}, T_{14} \} ]$

\vspace{5pt}\noindent
\textbf{\textit{Step 3}}
\begin{table}[htbp]
    \centering
    \begin{tabular}{|c|c|}
        \hline
        \textbf{TTS in the list} & \textbf{New cycles to add} \\
        \hline
        $\{ T_0 \}$ & - \\
        \hline
        $ \{ T_0 \} \cup s(C_1) \texttt{=} \{ T_0, T_1, T_3, T_4, T_5, T_6, T_7 \}$ & - \\
        \hline
        $ \{ T_0 \} \cup s(C_4) \texttt{=} \{ T_0, T_1, T_3, T_4, T_5, T_7, T_{10}, T_{11}, T_{14} \}$ & - \\
        \hline 
        $ \{ T_0 \} \cup s(C_1) \cup s(C_3) \texttt{=} \{ T_0, T_1, T_3, T_4, T_5, T_6, T_7, T_8, T_9 \} $ & $C_4$ \\
        \hline 
        $ \{ T_0 \} \cup s(C_1) \cup s(C_4) \texttt{=} \{ T_0, T_1, T_3, T_4, T_5, T_6, T_7, T_{10}, T_{11}, T_{14} \}$ & $C_2$ \\
        \hline
        $ \{ T_0 \} \cup s(C_4) \cup s(C_2) \texttt{=} \{ T_0, T_1, T_3, T_4, T_5, T_7, T_{10}, T_{11}, T_{12}, T_{13}, T_{14} \}$ & $C_3$ \\
        \hline
        $ \{ T_0 \} \cup s(C_4) \cup s(C_3) \texttt{=} \{ T_0, T_1, T_3, T_4, T_5, T_7, T_8, T_9, T_{10}, T_{11}, T_{14} \}$ & - \\
        \hline
    \end{tabular}
    \label{tab:step3}
\end{table} \par
Resulting TTS list: $[ \{ T_0 \}, \{ T_0, T_1, T_3, T_4, T_5, T_6, T_7 \}, \{ T_0, T_1, T_3, T_4, T_5, T_7, T_{10}, T_{11}, T_{14} \}, \\ 
\{ T_0, T_1, T_3, T_4, T_5, T_6, T_7, T_8, T_9 \}, \{ T_0, T_1, T_3, T_4, T_5, T_6, T_7, T_{10}, T_{11}, T_{14} \}, \\\{ T_0, T_1, T_3, T_4, T_5, T_7, T_{10}, T_{11}, T_{12}, T_{13}, T_{14} \}, \{ T_0, T_1, T_3, T_4, T_5, T_7, T_8, T_9, T_{10}, T_{11}, T_{14} \}, \\  \{ T_0, T_1, T_3, T_4, T_5, T_6, T_7, T_8, T_9, T_{10}, T_{11}, T_{14} \},  \{ T_0, T_1, T_3, T_4, T_5, T_6, T_7, T_{10}, T_{11}, T_{12}, T_{13}, T_{14} \}, \\ \{ T_0, T_1, T_3, T_4, T_5, T_7, T_8, T_9, T_{10}, T_{11}, T_{12}, T_{13}, T_{14} \} ]$ 

\vspace{5pt}\noindent
\textbf{\textit{Step 4}}
\begin{table}[htbp]
    \centering
    \begin{tabular}{|c|c|}
        \hline
        \textbf{TTS in the list} & \textbf{New cycles to add} \\
        \hline
        $\{ T_0 \}$ & - \\
        \hline
        $ \{ T_0 \} \cup s(C_1) \texttt{=} \{ T_0, T_1, T_3, T_4, T_5, T_6, T_7 \}$ & - \\
        \hline
        $ \{ T_0 \} \cup s(C_4) \texttt{=} \{ T_0, T_1, T_3, T_4, T_5, T_7, T_{10}, T_{11}, T_{14} \}$ & - \\
        \hline 
        $ \{ T_0 \} \cup s(C_1) \cup s(C_3) \texttt{=} \{ T_0, T_1, T_3, T_4, T_5, T_6, T_7, T_8, T_9 \} $ & - \\
        \hline 
        $ \{ T_0 \} \cup s(C_1) \cup s(C_4) \texttt{=} \{ T_0, T_1, T_3, T_4, T_5, T_6, T_7, T_{10}, T_{11}, T_{14} \}$ & - \\
        \hline
        $ \{ T_0 \} \cup s(C_4) \cup s(C_2) \texttt{=} \{ T_0, T_1, T_3, T_4, T_5, T_7, T_{10}, T_{11}, T_{12}, T_{13}, T_{14} \}$ & - \\
        \hline
        $ \{ T_0 \} \cup s(C_4) \cup s(C_3) \texttt{=} \{ T_0, T_1, T_3, T_4, T_5, T_7, T_8, T_9, T_{10}, T_{11}, T_{14} \}$ & - \\
        \hline 
        $\{ T_0 \} \cup s(C_1) \cup s(C_3) \cup s(C_4)\texttt{=} \{ T_0, T_1, T_3, T_4, T_5, T_6, T_7, T_8, T_9, T_{10}, T_{11}, T_{14} \}$ & $C_2$ \\
        \hline 
        $\{ T_0 \} \cup s(C_1) \cup s(C_4) \cup s(C_2)\texttt{=} \{ T_0, T_1, T_3, T_4, T_5, T_6, T_7, T_{10}, T_{11}, T_{12}, T_{13}, T_{14} \}$ & - \\
        \hline 
        $\{ T_0 \} \cup s(C_4) \cup s(C_2) \cup s(C_3) \texttt{=} \{ T_0, T_1, T_3, T_4, T_5, T_7, T_8, T_9, T_{10}, T_{11}, T_{12}, T_{13}, T_{14} \}$ & - \\
        \hline
    \end{tabular}
    \label{tab:step4}
\end{table} \par
Resulting TTS list: $[ \{ T_0 \}, \{ T_0, T_1, T_3, T_4, T_5, T_6, T_7 \}, \{ T_0, T_1, T_3, T_4, T_5, T_7, T_{10}, T_{11}, T_{14} \}, \\ 
\{ T_0, T_1, T_3, T_4, T_5, T_6, T_7, T_8, T_9 \}, \{ T_0, T_1, T_3, T_4, T_5, T_6, T_7, T_{10}, T_{11}, T_{14} \}, \\\{ T_0, T_1, T_3, T_4, T_5, T_7, T_{10}, T_{11}, T_{12}, T_{13}, T_{14} \}, \{ T_0, T_1, T_3, T_4, T_5, T_7, T_8, T_9, T_{10}, T_{11}, T_{14} \}, \\  \{ T_0, T_1, T_3, T_4, T_5, T_6, T_7, T_8, T_9, T_{10}, T_{11}, T_{14} \},  \{ T_0, T_1, T_3, T_4, T_5, T_6, T_7, T_{10}, T_{11}, T_{12}, T_{13}, T_{14} \}, \\ \{ T_0, T_1, T_3, T_4, T_5, T_7, T_8, T_9, T_{10}, T_{11}, T_{12}, T_{13}, T_{14} \}, \\\{ T_0, T_1, T_3, T_4, T_5, T_6, T_7, T_8, T_9, T_{10}, T_{11}, T_{12}, T_{13}, T_{14} \} ]$
\par
After the insertion of $\{ T_0 \} \cup s(C_1) \cup s(C_3) \cup s(C_4) \cup s(C_2)$ in the list, no further cycle can be added, and the algorithm terminates.

\begin{algorithm}
\caption{Find Seed Trace Transitions Sets for Simple Paths from a Node to Another} 
\label{algo:find_simple_paths}
\KwIn{Graph $G$, First node $node_1$, Second node $node_2$, Current path $currentPath$, List of paths $paths$}
\KwOut{List of trace transitions sets for simple paths from the first node to the second node}

\SetAlgoLined
\SetKwFunction{FMain}{FindSimplePaths}
\SetKwProg{Fn}{Function}{:}{end}

\Fn{\FMain{$G$, $currentNode$, $endNode$, $currentPath$, $paths$}}{
    \If{$currentNode$ is the same as $ endNode$}{
        Append $currentPath$ to $paths$ \;
    }
    \Else{
        \ForEach{$edge$ \textbf{in} $G$}{
            \If{$edge$ is directed from $node_1$ to $otherEnd$}{
                FindSimplePaths($graph$, $ otherEnd$, $node_2$, $currentPath + [edge]$, $paths$) \;
            }
        }
    }
    \Return{} the list of sets of edges corresponding to $paths$ (set of trace transitions sets) \;
}

\end{algorithm}

\begin{algorithm}
\caption{Find All Trace Transitions Sets given a List of seed Trace Transitions Sets} 
\label{algo:find_complex_paths}
\KwIn{list of seed trace transitions sets $pathSetsList$, set of cycles $cycles$}
\KwOut{output added to $pathSetsList$}

\SetAlgoLined
\SetKwFunction{FMain}{FindComplexPaths}
\SetKwProg{Fn}{Function}{:}{end}

\Fn{\FMain{$pathSetsList$, $cycles$}}{
    \ForEach{$pathSet$ in $pathSetsList$}{
        \ForEach{$edge$ in $pathSet$}{
            Let $edge$ be directed from $startNode$ to $endNode$ \;
            \ForEach{$cycle$ in $cycles$}{
                \If{\text{an edge exists in} $cycle$ \text{starting from} $startNode$ or $endNode$}{
                    $newPathSet \gets pathSet \cup cycle$ \;
                    \If{$newPathSet$ is not in $pathSetsList$}{
                        Insert $newPathSet$ in $pathSetsList$ \;
                    }
                }
            }
        }
    }
}

\end{algorithm}

\subsection*{Obtaining the Equivalence Mapping}
We use the two algorithms presented above to lead to equivalence mappings for each marking in the old WF-net. The procedure is outlined in Algorithm \ref{algo:find_equivalence_mapping}. We compute the list of TTSs for each node in the old and new reachability graphs using algorithms \ref{algo:find_simple_paths} and \ref{algo:find_complex_paths}. For each TTS, we find the nodes in the new reachability graph that have the same TTS. These nodes are history equivalent to the corresponding node in the old graph.

\subsection*{Empty Transitions}
\begin{figure}[htbp]
    \centering
    \includegraphics[width=0.6\textwidth]{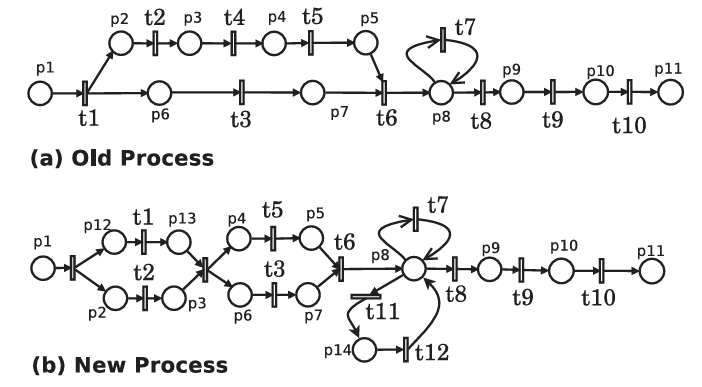}
    \caption{An example taken from \emph{A Taxonomy of Consistency Models in Dynamic Migration of Business Processes} \cite{a_taxonomy}}
    \label{fig:empty_transitions}
\end{figure}

A WF-net may also have empty helper transitions as given in the new process in Figure \ref{fig:empty_transitions}. Such empty transitions are often used for maneuvering process logics as additional helper transitions that do not correspond to any task in the workflow process, but are needed only to implement synchronization points such as AND forks and joins. To consider empty transitions in the net, we modify the algorithm to first remove all empty transitions from the TTSs since these transitions will have no effect while matching compatible TTSs. While comparing TTSs, the empty transitions are ignored. The set of nodes found thus defines the equivalent mapping for a given node. This procedure is repeated for all nodes in the old net's reachability graph.

As a snapshot example, the equivalence mapping found for the old and new WF-nets given in Figure \ref{fig:empty_transitions} is given below. The remaining markings in the old WF-net have no equivalent states in the new net, and therefore they are in the change region.
The complete algorithm is outlined in Algorithm \ref{algo:find_equivalence_mapping}. For the old and new nets in Figure \ref{fig:empty_transitions}, the algorithm generates the equivalence-mapping table for the migrated WF-net as shown in Table \ref{tab:mapping}.
  
\begin{table}[htbp]
    \centering
    \begin{tabular}{|c|c|}
        \hline
        Old Net Marking & New Net Equivalent Markings \\
        \hline
        $\{ \text{p}_3, \text{p}_7 \}$ & $\{ \text{p}_4, \text{p}_7 \}$ \\
         $\{ \text{p}_3, \text{p}_6 \}$ & $\{ \text{p}_{13}, \text{p}_3 \}, \{ \text{p}_4, \text{p}_6 \}$ \\
        $\{ \text{p}_2, \text{p}_6 \}$ & $\{ \text{p}_{13}, \text{p}_2 \}$ \\
        $\{ \text{p}_1 \}$ & $\{\text{p}_1 \}, \{ \text{p}_{12}, \text{p}_2 \}$ \\
        $\{ \text{p}_{2}, \text{p}_7 \}$ & $\phi$ (change region)\\
        $\{ \text{p}_{4}, \text{p}_6 \}$ & $\phi$ (change region)\\
        $\{ \text{p}_{4}, \text{p}_7 \}$ & $\phi$ (change region)\\
        $\{ \text{p}_{5}, \text{p}_6 \}$ & $\phi$ (change region)\\
        $\{ \text{p}_{5}, \text{p}_7 \}$ & $\phi$ (change region)\\
        $\{ \text{p}_{8} \}$ & $\phi$ (change region)\\
        $\{ \text{p}_{9} \}$ & $\phi$ (change region)\\
        $\{ \text{p}_{10} \}$ & $\phi$ (change region)\\
        $\{ \text{p}_{11} \}$ & $\phi$ (change region)\\
        \hline
    \end{tabular}
    \caption{Equivalence mapping for the old and new WF-nets in Fig. \ref{fig:empty_transitions}}
    \label{tab:mapping}
\end{table}

\begin{figure}
    \centering
    \includegraphics[width=\textwidth]{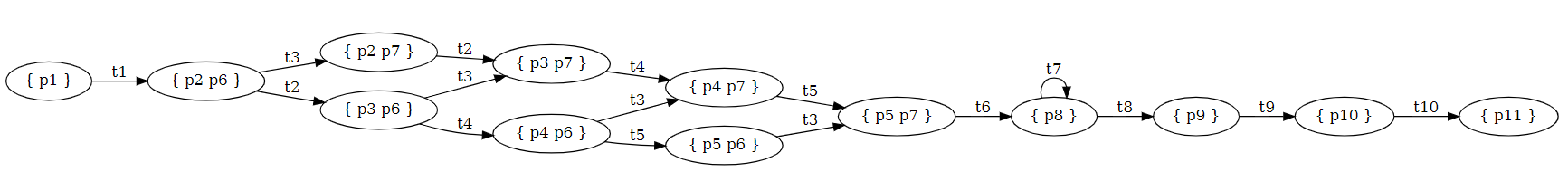}
    \caption{Reachability for the old process from Figure \ref{fig:empty_transitions}}
    \label{fig:old8_rg}
\end{figure}

\begin{figure}[htbp]
    \centering
    \includegraphics[width=\textwidth]{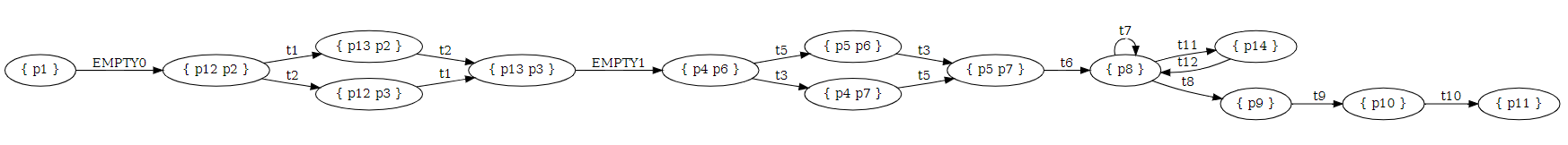}
    \caption{Reachability for the new process from Figure \ref{fig:empty_transitions}}
    \label{fig:new8_rg}
\end{figure}

\begin{algorithm}
\caption{Find Equivalence Mapping}
\label{algo:find_equivalence_mapping}
\KwIn{Old Workflow net $W_{\text{old}}$, New Workflow net $W_{\text{new}}$}
\KwOut{Equivalence mapping table}

\SetAlgoLined
\SetKwFunction{FMain}{FindEquivalenceMapping}
\SetKwProg{Fn}{Function}{:}{end}

\Fn{\FMain{$W_{\text{old}}, W_{\text{new}}$}}{
    Initialize an empty mapping table $mappingTable$ \;
    
    Initialize an empty map $S_{new}$ to store the list of trace transitions sets for all states in the reachability graph of $W_{\text{new}}$ \;

     \ForEach{state $N$ in the reachability graph of $W_{\text{new}}$}{
        $S_N \gets $ set of TTSs leading to $N$ (obtained using Algorithms \ref{algo:find_simple_paths} and \ref{algo:find_complex_paths}) \;
        Let $emptyTransitions_{\text{new}}$ be the set of empty transitions present in $W_{\text{new}}$ \;
        $S_{new}[N] \gets \{ tts \setminus emptyTransitions_{\text{new}} \hspace{1mm} | \hspace{1mm} tts \in S_N \}$  \;
     }
    
    \ForEach{state $N$ in the reachability graph of $W_{\text{old}}$}{
        $S_N \gets $ set of TTSs leading to $N$ (obtained using Algorithms \ref{algo:find_simple_paths} and \ref{algo:find_complex_paths})  \;
        Let $emptyTransitions_{\text{old}}$ be the set of empty transitions present in $W_{\text{old}}$  \;
        $S_N^\prime \gets \{ tts \setminus emptyTransitions_{\text{old}} \hspace{1mm} | \hspace{1mm} tts \in S_N \}$  \;
        Initialize an empty list $mappingTable[N]$ \;
        \ForEach{$N_{new}$ in $S_{new}$}{
            \If{$S_{new}[N_{new}] \cap S_N^\prime \neq \phi$}{
                Append $N_{new}$ to $mappingTable[N]$ \;
            }
        }
       
    }
    
    \Return{$mappingTable$} \;
}

\end{algorithm}

\section{Intuition Behind the Algorithm} \label{intuition}


\noindent
\textbf{Claim 1}: We obtain the exhaustive list of TTSs for a node using Algorithms \ref{algo:find_simple_paths} and \ref{algo:find_complex_paths}
\par \noindent
\textbf{Proof}: Obtaining non-cyclic paths is a straightforward task. Every path that involves some cycles can be broken down into two components - the cycles and the seed path without any cycle. Since the reachability graphs are finite, we only have finitely many cycles in the graph, all obtained using DFS. Let $C$ be the set of all cycles in the graph, and let $C_p$ denote the set of cycles with some vertex or edge in common with a path $p$. Then we can also traverse all $c \in C_p$ while traversing $p$. The set of all possible TTSs that can be obtained from a seed TTS $p$ is, therefore, given by $P = \{ p \cup s(c) \hspace{1mm}|\hspace{1mm} p \in P, \hspace{1mm} c \in C_p\}$, where $s(c)$ denotes the set of edges in the cycle $c$. Our algorithm recursively computes this set $P$ for all paths by starting from the acyclic seed paths, till no further path can be added, thereby obtaining the exhaustive list of TTSs for all possible paths through which one can reach each node in the graph.

\par
\noindent
\textbf{Claim 2}: The main algorithm given in Algorithm \ref{algo:find_equivalence_mapping} generates a history equivalence-based mapping for each node. \\
\textbf{Proof}: Let $S_{n\_old}$ be the set of all TTSs in the old net leading to a state $n_{old}$, and $S_{n\_new}$ be the set of all TTSs in the new net leading to a state $n_{new}$. $n_{old}$ is equivalent to $n_{new}$ if $S_{n\_old} \cap S_{n\_new} \neq \phi$. We compute $S_{n\_old} \cap S_{n\_new}$ for each pair $(n_{old}, n_{new})$, so for any two nodes that have a TTS in common, the node from the new graph is marked as equivalent to the node from the old graph. \par
Claim 1 proves that $S_{n\_old}$ 
and $S_{n\_new}$ are exhaustive sets of TTSs. This completes the proof of correctness of Algorithm \ref{algo:find_equivalence_mapping}.


\section{Conclusion}
In this paper, we have proposed an algorithm that gives an equivalence mapping for processes that undergo dynamic migration. We use the notion of history equivalence as the criterion for identifying equivalent states, in which, history is represented by the set of all possible trace transition sets. If one of them is found to be common between the old and the new nets for a given state in the old net, the corresponding state found in the new net becomes its history equivalent migration point. As future work, we can further extend the algorithm to consider other trail-based consistency models like trace equivalence and purged history equivalence.

\bibliographystyle{IEEEtran}
\bibliography{main}

\end{document}